\documentclass[aps,prl,twocolumn,groupedaddress]{revtex4}
\usepackage[dvips]{graphicx}
\usepackage{amssymb}
\usepackage{color}
\newcommand{\T}{{\cal T}}

\newcommand{\V}{{\cal V}}

\begin{document}

\title{Noise dephasing in the edge states of the Integer Quantum Hall regime}

\author{P. Roulleau, F. Portier, and P. Roche}
\affiliation{CEA Saclay , Service de Physique de l'Etat
Condens{\'e}, Nanoelectronic group, F-91191 Gif-sur-Yvette,
France}
\author{A. Cavanna, G. Faini, U. Gennser, and D. Mailly}
\affiliation{CNRS, Laboratoire de Photonique et Nanostructures,
Phynano team, Route de Nozay, F-91460 Marcoussis, France}

\date{\today}

\begin{abstract}
An electronic Mach Zehnder interferometer is used in the integer
quantum hall regime at filling factor 2, to study the dephasing of
the interferences. This is found to be induced by the electrical
noise existing in the edge states capacitively coupled to each
others. Electrical shot noise created in one channel leads to
phase randomization in the other, which destroys the interference
pattern. These findings are extended to the dephasing induced by
thermal noise instead of shot noise: it explains the underlying
mechanism responsible for the finite temperature coherence time
$\tau_\varphi(T)$ of the edge states at filling factor $2$,
measured in a recent experiment. Finally, we present here a theory
of the dephasing based on Gaussian noise, which is found in
excellent agreement with our experimental results.
\end{abstract}

 \maketitle

Although many experiments in quantum optics can be reproduced with
electron beams using the edge states of the Integer Quantum Hall
Effect (IQHE), there exist fundamental differences due to the
Coulomb interaction. As an example, the Mach-Zenhder type of
interferometer in the IQHE \cite{Ji03Nature422p415} has recently
allowed to observe quantum interferences with the unprecedented
 90\% visibility \cite{Neder07Nature448p333}, opening a new field of promising
quantum information experiments. Indeed, the edge states of the
IQHE provide a way to obtain 'ideal' uni-dimensional quantum
wires. However, very little is known about the decoherence
processes in these 'ideal' wires. Only very recently their
coherence length was quantitatively determined as well as its
temperature dependence established \cite{Roulleau0710.2806}. Here,
we show that the underlying mechanism responsible for the finite
coherence length is the thermal noise combined with the poor
screening in the IQHE regime \cite{Seelig01PRB64n245313}.

In the IQHE, gapless excitations develop on the edge of the sample
and form one dimensional chiral wires (edge states), the number of
which is determined by the number of electrons per quantum of flux
(the filling factor $\nu$). In these wires, the electrons drift
along the edge in a beam-like motion making experiments usually
done with photons possible with electrons. The choice of the
filling factor at which one obtains high visibility interferences
requires a compromise between a magnetic field high enough to form
well defined edge states, and small enough to still deal with a
good Fermi liquid. Na{\"i}vely one could think that the highest
visibility would have been observed at $\nu=1$, but it is not
actually the case\cite{Ji03Nature422p415}. This is most probably
due to decoherence induced by low energy collective spin
excitations (skyrmions \cite{Barrett95PRL74p5112}) making spin
flip processes possible. In practice, the highest visibility (90\%
\cite{Neder07Nature448p333}) has been obtained at filling factor
2, when there are two spin polarized edge states. Here, chirality
and uni-dimensionality prevent first order inelastic scattering in
the wires themselves \cite{Martin90PRL64p1971}, while tunneling
from one edge to the other requires spin flip
\cite{Dixon97PRB56p4743}.

To show that the origin of the finite coherence length is related
to the coupling between two neighbouring edge states, we have
proceeded as follow. First we have made a which-path experiment
inducing on purpose shot noise on the inner channel while
measuring the outer channel interferences. The visibility decrease
is shown to result from a gaussian noise, in opposition to a
recent experiment \cite{Neder07NatureP3p534}. Using the parameters
extracted from the which-path measurements, we are able to
calculate the dephasing resulting from thermal noise (instead of
shot noise). The result is in perfect agreement with our recent
measurements of the finite temperature coherence length
\cite{Roulleau07PRB76n161309}. Moreover, the magnetic field
dependence of the coherence length is shown to result from a
variation of the coupling between the two edges. Finally, we have
developed a theory which gives a full scheme of the dephasing
mediated by the electronic noise.

\begin{figure}[t]
\centerline{\includegraphics[angle=-90,width=8cm,keepaspectratio,clip]{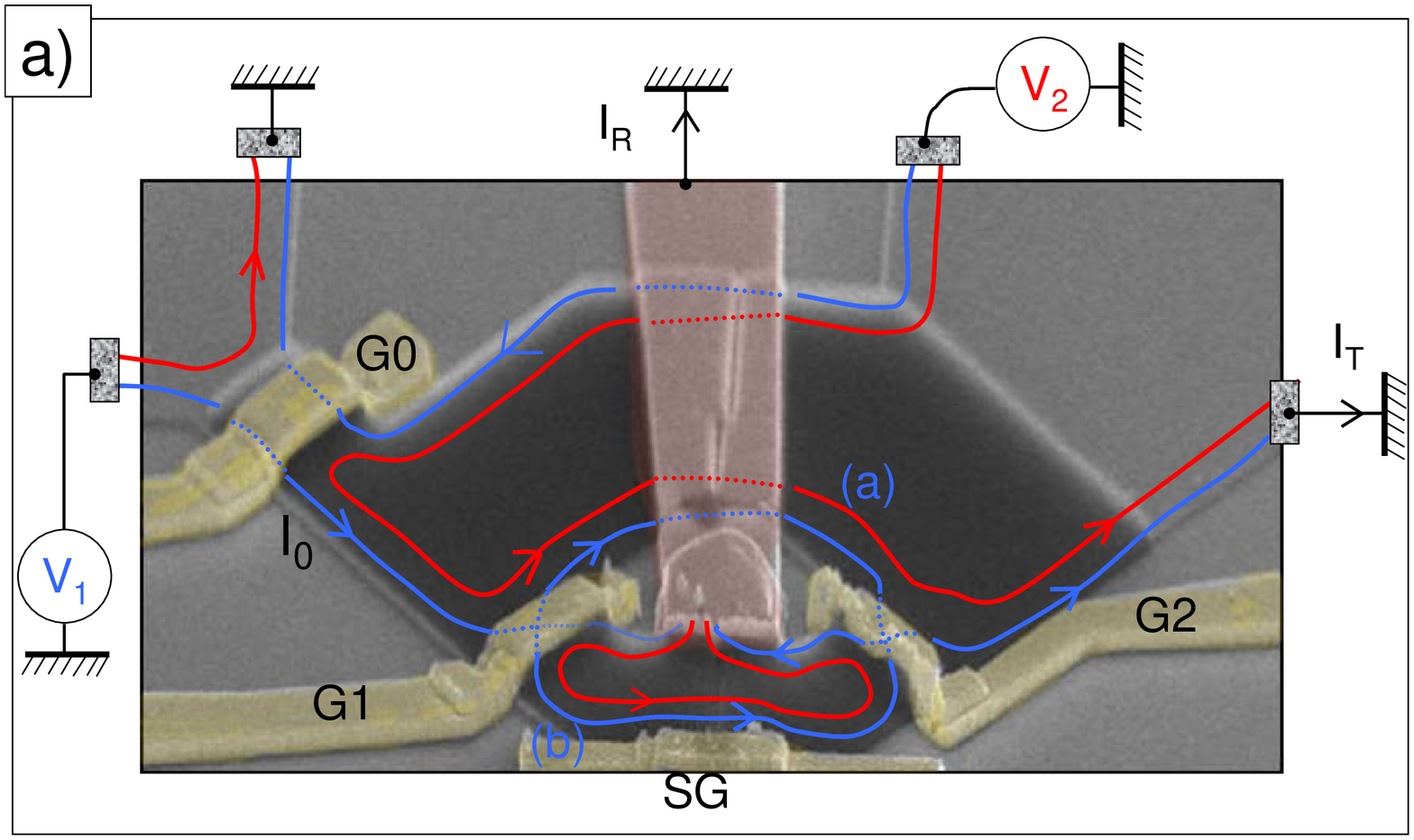}}
\centerline{\includegraphics[angle=-90,width=8cm,keepaspectratio,clip]{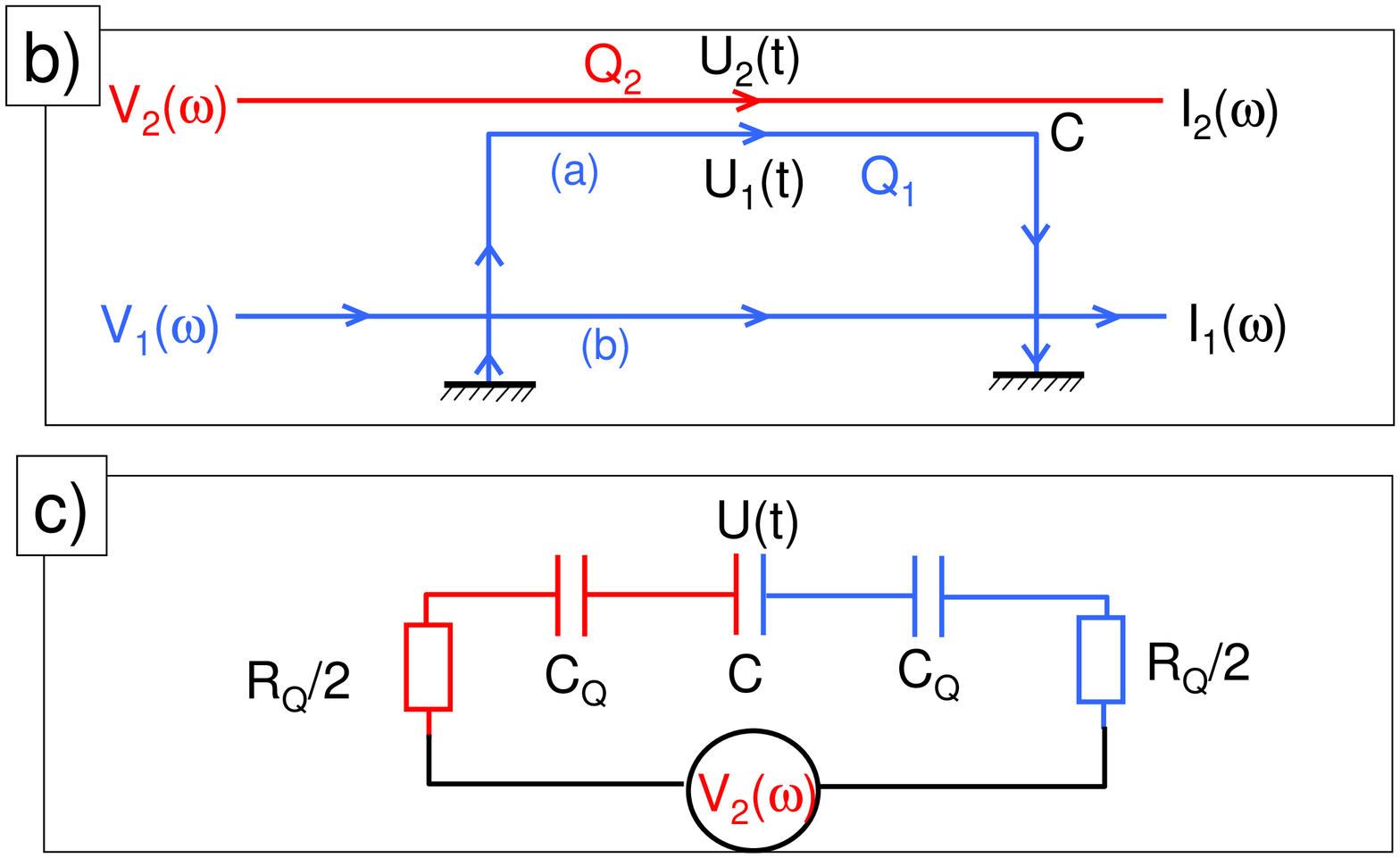}}
\caption{(color online) \textbf{a)}Tilted SEM view of the device,
with schematic representation of the edge states. G1 and G2 are
Quantum Point Contact (QPC) which define the two beam splitters of
the Mach-Zehnder interferometer. They are set to transmission
$\T_1\sim\T_2 \sim 1/2$ for the outer channel, while fully
reflecting the inner channel. The two arms (a) and (b) are
$L$=11.3~$\mu$m long defining a surface $S$ of 34~$\mu$m$^2$. The
small inner ohmic contact is connected to the ground via an Au
metallic bridge. SG is a side gate. G0 is an additional beam
splitter which makes it possible to bias the inner edge state by
$V_2$, while the other is biased by $V_1$. G0 is tuned such that
the outer channel is fully reflected, while the inner is
transmitted with a probability $\T_0$. \textbf{b)} Schematic
representation of the edge states coupled by a geometrical
capacitance $C$. \textbf{c)} Low frequency equivalent circuit with
$V_1$ set to 0~V. $V_2(\omega)$ can be either generated by shot
noise or by thermal noise, $C_Q=\tau/R_Q$ and $R_Q=h/e^2$ (see
text).}{\label{figure1}}
\end{figure}

The interferences are obtained using an electronic Mach-Zehnder
Interferometer (MZI) which was patterned on a high mobility two
dimensional electron gas at a GaAs/Ga$_{1-x}$Al$_x$As
heterojunction (density $n_S=2.0\times10^{11}$cm$^{-2}$ and
mobility $\mu=2.5\times10^6$cm$^2$/Vs). Measurements have been
done in the quantum Hall regime, at filling factor 2 (with a
magnetic field B=5.2 T). In the edge states, the electrons have a
chiral motion with a drift velocity of the order of
$10^4-10^5$~ms$^{-1}$. A SEM view of the sample as well as a
schematic representation of the two edge states are shown in Fig.
1a. The outer incoming edge state is split by G1 in two paths (a)
and (b), which are recombined at G2 leading to interferences. SG
is a side gate used to change the surface $S$ defined by the two
arms of the interferometer. The current which is not transmitted
through the MZ, I$_R$=I$_0$$-$I$_T$, is collected to the ground
via the inner ohmic contact.  The differential transmission
$\T=dI_t/dI_0$ have been measured at low temperature ( $\sim$
20~mK) by standard lock-in techniques with an AC voltage ($V_1\sim
1\mu$V$_{RMS}$ at 619 Hz).

It is straightforward to show that $\T\propto[1+\V\sin(\varphi)]$,
$\V$ being the visibility and $\varphi$ the Aharonov-Bohm (AB)
flux through the surface $S$ of the MZI
\cite{Roulleau07PRB76n161309}. In the present study we tuned the
transmission $\T_1$ and $\T_2$ of the beam splitters G1 and G2 to
1/2 in order to have a maximum visibility. The interferences are
revealed by varying $\varphi$. It can be done either by applying a
voltage $V_{SG}$ on the side gate, or by applying a voltage $V_2$
on the inner edge state (playing here a role similar to the side
gate). In Fig. 2, we have plotted the interference pattern
obtained by the two methods. The periodicity $V_0$ of
interferences with respect to $V_2$ depends on the coupling
between the two edge states which will be shown to be related to
the time of flight $\tau$ through the MZI. In Fig. 4, one can
notice that $V_0$ exhibits a large (up to a factor 3) non
monotonous variation with the magnetic field on the Hall plateau
at $\nu=2$.

Any fluctuations on $V_2$ blur the phase. For a Gaussian
distribution of the phase (we will discuss this notion later), the
visibility is proportional to $e^{-<\delta\varphi^2>/2}$
\cite{Stern90PRA41p3436} where $<\delta\varphi^2>$ is the variance
of the Gaussian distribution. It is simply related to the noise
power spectrum $S_{22}$ of $V_2$ through the coupling constant and
the (unknown) bandwidth $\Delta\nu$: $ <\delta\varphi^2> =
(2\pi)^2 <\delta V_2^2>/V_0^2 = (2\pi)^2 S_{22}\Delta\nu/V_0^2$.
If one generates partition noise on the inner edge state tanks to
the splitter $G_0$, the resulting excess noise $\Delta S_{22}=
2eR_Q\T_0(1-\T_0)V_2[\coth(eV_2/(2k_BT))-2k_BT/(eV)]$
\cite{Martin92PRB45p1742,Buttiker92PRB46p12485} leads to a
visibility decreasing exponentially with $V_2$ when $eV_2\gg
k_BT$:

\begin{equation}
\V =\V_0 e^{-\T_0(1-\T_0)(V_2-2k_BT/e)/V_\varphi},
\end{equation} with \begin{equation} V_\varphi^{-1} =
\frac{4\pi^2eR_Q}{V_0^2}\Delta \nu, \label{equationVphi}
\end{equation}
and $R_Q=1/G_Q=h/e^2$.
\begin{figure}[h]
\centerline{\includegraphics[angle=-90,width=8.5cm,keepaspectratio,clip]{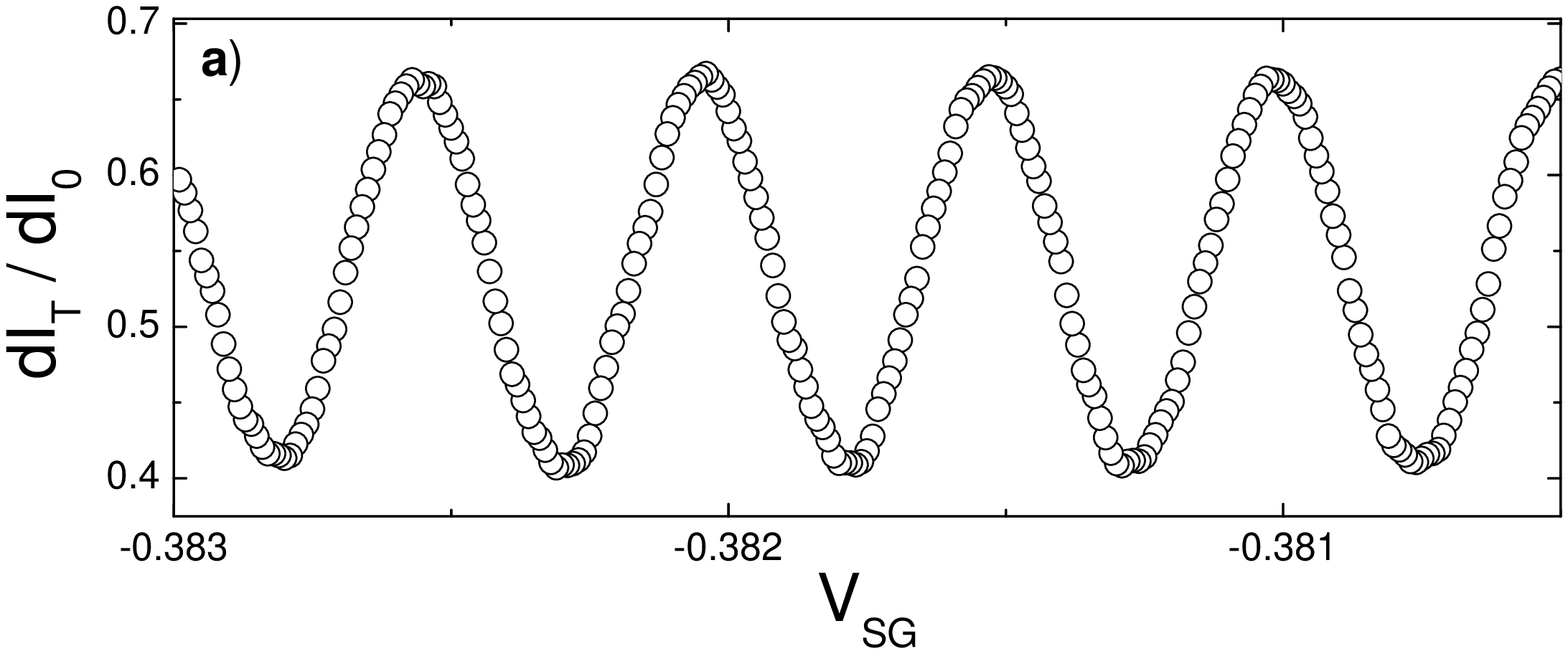}}
\centerline{\includegraphics[angle=-90,width=8.5cm,keepaspectratio,clip]{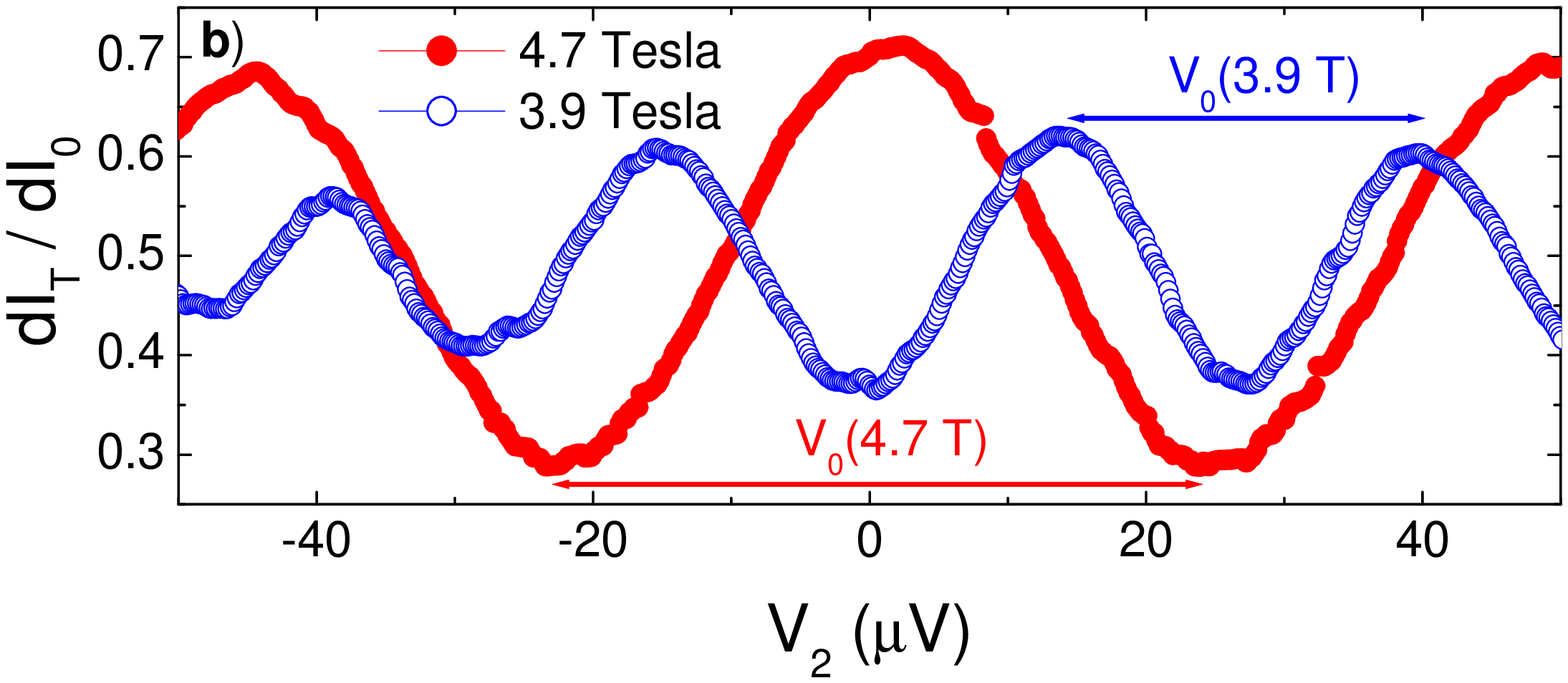}}
\caption{(color online) \textbf{a)} Phase sweeping by varying the
side gate voltage $V_{SG}$. \textbf{b)} Phase sweeping of the
interferometer upon varying $V_2$ with $\T_0=1$ for two different
magnetic fields. The periodicity $V_0$ of interferences  depends
on the magnetic field as shown in Fig. 4.}{\label{figure2}}
\end{figure}

In Eq. 1, the unknown parameter is $V_\varphi$ which is related to
the bandwidth $\Delta\nu$ (Eq. 2). This approach for the dephasing
is valid only if $\Delta\nu$ is such that the fluctuations lead to
a Gaussian distribution of $\varphi$. It implies that many
electrons have to be involved in the dephasing during the
measuring time $1/\Delta\nu$, namely that max($eV_2,2k_BT$)$\gg
h\Delta\nu$. This condition coincides with the fact that the noise
power spectrum $S_{22}$ can be considered as frequency
independent. Note that the dephasing rate increases with $V_2$
because the number of involved electrons increases, not because
the coupling between electrons increases with $V_2$ (as claimed in
\cite{Neder07NatureP3p534}). Fig. 3 shows that our data are in
remarkable agreement with Eq. \ref{equationVphi}. In Fig. 3a we
have plotted the visibility versus $V_2$ when $\T_0=1/2$, for two
different magnetic fields. $\V$ decreases exponentially with
$V_2$. The solid lines are fits to the data using an electronic
temperature of $25$~mK (for a fridge temperature of 20~mK).
$V_\varphi$ and $\V_0$ are the fitting parameters. The values of
$V_\varphi$ deduced from these measurements depend on the magnetic
field in the same way as $V_0$. In fact, $V_\varphi$ is found to
be proportional to $V_0$ (see Fig. 4). The slope of the
exponential decrease is modified by the transmission of the beam
splitter following a $\T_0(1-\T_0)$ law. Fig. 3b shows the
visibility for different values of $V_2$ and $\T_0$ at a magnetic
field of 4.6 Tesla. The solid lines are fits to the data using
Eq.1 with $V_\varphi=7.2$~$\mu$V and $T=25$~mK. Clearly, at high
bias there is no V-shape contrary to what has been recently
observed in ref.\cite{Neder07NatureP3p534}. Instead, the curves
show that the Gaussian approximation is valid. Note that the
agreement with our theory is perfect when $\T_0$ is well defined
in our sample. The dispersed data on the edges in Fig. 3b coincide
to a strong dependence of $\T_0$ with the voltage applied on G0,
resulting on an energy dependent transmission $\T_0$
\cite{Neder07NatureP3p534}.

We now compare the exponential decrease of the visibility in
presence of shot noise with our recent observation that the
coherence length of edge states is inversely proportional to the
temperature \cite{Roulleau0710.2806}. When $eV_2\ll k_BT$, the
noise is dominated by the Johnson-Nyquist noise $S_{22}=4k_BTR_Q$.
One obtains:

\begin{equation}
\V =\V_0 e^{-T/T_\varphi}\; \mathrm{with} \; T_\varphi^{-1} =
\frac{2\times8\pi^2k_BR_Q}{V_0^2}\Delta \nu. \label{equationTphi}
\end{equation}
Here, the factor 2 arises from the fact that the two arms of the
interferometer suffer from a coupling with a noisy inner channel,
instead of one when creating partitioning. From equations
\ref{equationVphi} and \ref{equationTphi}, one gets:

\begin{equation}
e V_{\varphi} = 4 k_B T_{\varphi}
\end{equation}

Fig.  \ref{expdecay.fig}, which is our main result, shows that Eq.
4 is in very good agreement with our data. This demonstrates for
the first time that thermal noise and coupling between the two
edge states are responsible for the finite coherence length
measured recently \cite{Roulleau0710.2806}.
\begin{figure}[h]
\centerline{\includegraphics[angle=-90,width=8.5cm,keepaspectratio,clip]{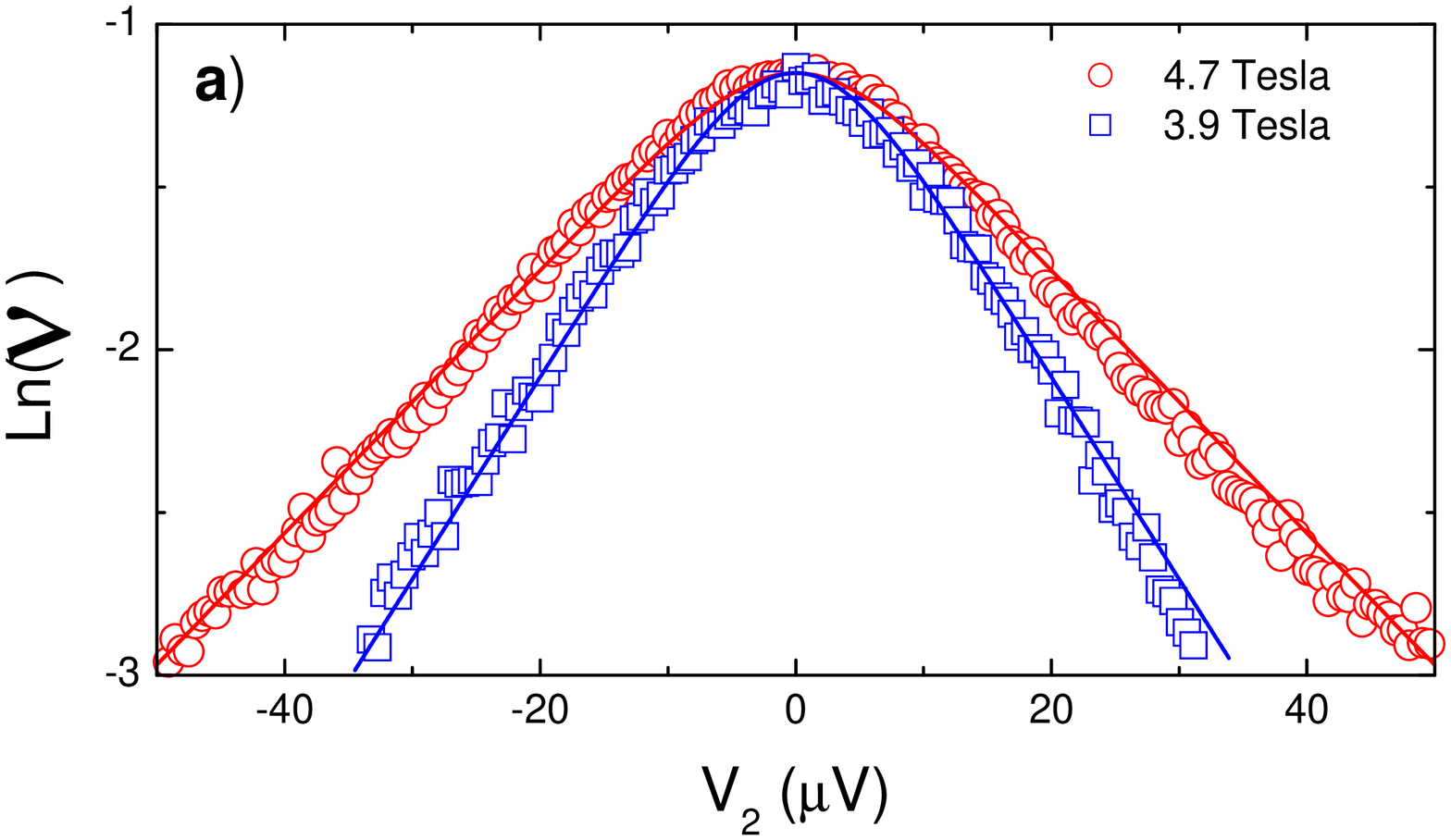}}
\centerline{\includegraphics[angle=-90,width=8.5cm,keepaspectratio,clip]{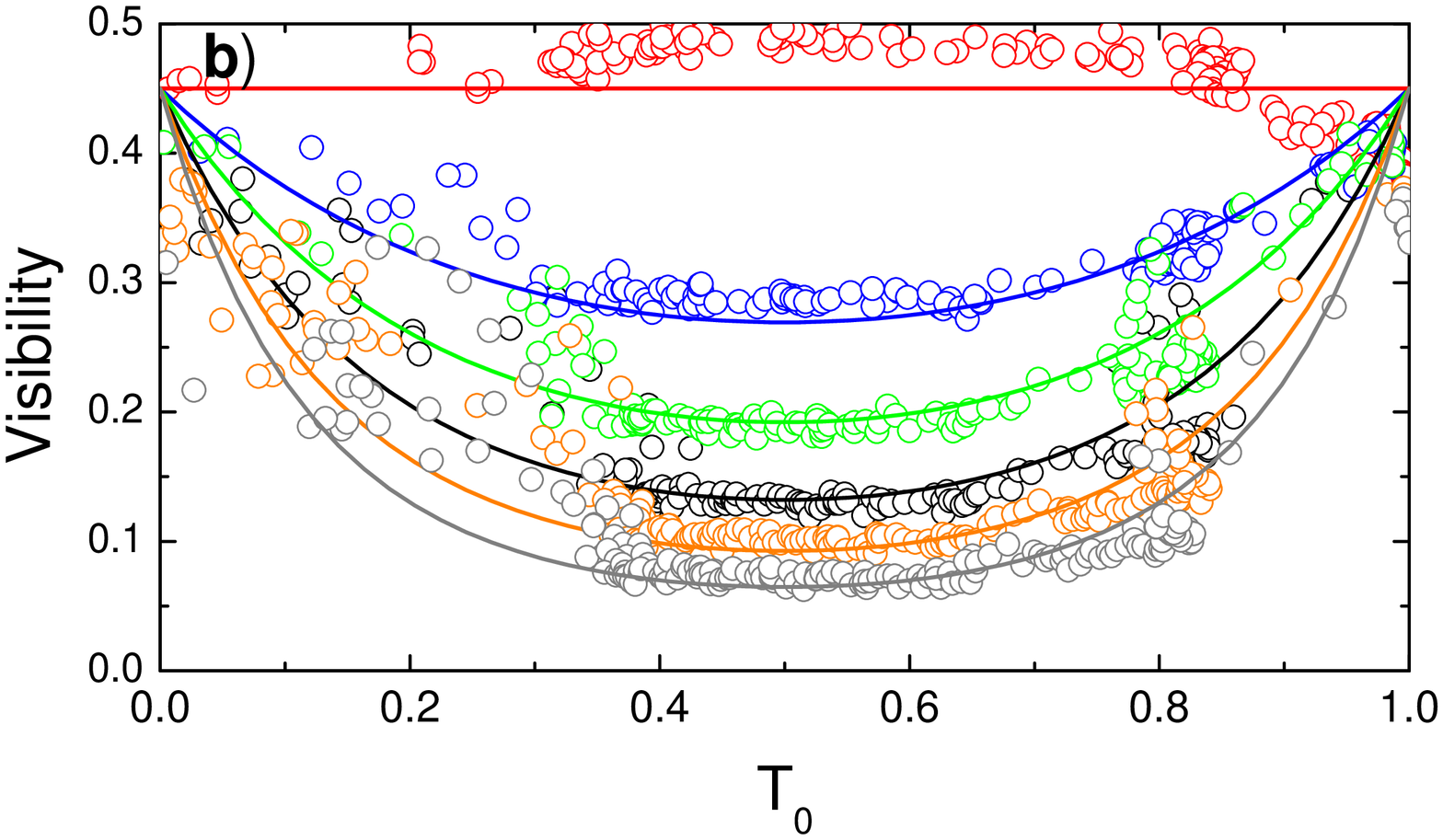}}
\caption{(color online) \textbf{a)} Visibility decrease of the
interferometer as a function of $V_2$ at $\T_0=1/2$ for two
different magnetic fields 4.7 and 3.9 T. The solid lines are fit
to the data $\V=\V_0e^{-2\pi^2\Delta S_{22}\Delta\nu/V_0^2}$ with
an electronic temperature of 25~mK (for a base temperature of
20~mK) and $\T_0=1/2$. The high bias fit of the exponential
decrease $\V=\V_0\exp(-\T_0(1-\T_0)V_2/V_\varphi)$ allows us to
determine $V_\varphi$ which is found to depends on the magnetic
field. \textbf{b)} Visibility decrease of the interferometer as a
function of $\T_0$ for $V_2$=0, 21, 31, 42, 53 and 63 $\mu$V from
top to bottom. The solid lines are fits to the data using Eq. 1
with $\V_0$=0.45, $V_\varphi = 7.2$~$\mu$V and
$T=25$~mK.}{\label{figure3}}
\end{figure}

This figure also brings a valuable point for the understanding of
the underlying physics: the proportionality  of $V_{\varphi}$ to
$V_0$. This may be surprising at first sight: if $\Delta \nu$ were
constant, according to Eqs. \ref{equationVphi} and
\ref{equationTphi}, $V_{\varphi}$ would scale as $V_0^2$. Instead,
as seen in Fig. \ref{expdecay.fig}, $V_{\varphi}$ is proportional
to $V_0$. We will show that, varying the magnetic field changes
the time of flight $\tau$ through the MZI, thus changing both the
coupling between the edge states and the bandwidth $\Delta \nu$.
This accounts for the proportionality of $V_0$ and $V_{\varphi}$.
From the measurements of $V_0$ and $V_\varphi$, one can deduce
using Eq. 2 that $h\Delta \nu$ varies from $\sim$3 to
$\sim$7~$\mu$eV when changing the magnetic field. This value of
$h\Delta\nu$ is $\lesssim2k_BT$, which validates our approach of
Gaussian and white noise \footnote{ It can lead to a small
deviation from our theory at the lowest temperature (20 mK) when
$eV_2\le 2k_BT$.}.

We now turn to a theoretical approach that relates $V_0$ and
$\Delta \nu$ to microscopic parameters of the system and explains
why $\Delta\nu\propto V_0$. We first consider the capacitive
coupling between edges states.  The influence of a gate coupled to
one arm of the interferometer can be modelled in two equivalent
ways.  One can either consider that a gate voltage changes the
surface $S$ of the MZI without modifying the potential $U_1$ felt
by the electrons, or that it adds an excess charge
\cite{Stern90PRA41p3436,Seelig01PRB64n245313,Rohrlich07PRL98n096803}
which modifies $U_1$ without changing $S$. In this picture, the
phase changes by $\varphi=\int_0^\tau eU_1\mathrm{d}t/\hbar$,
where $\tau = L/v_D$ is the time of flight through the MZI and $L$
stands for the length of one arm of the interferometer. This last
relation allows to relate the phase noise $S_\varphi(\omega)$ to
the potential noise $S_{U_1U_1}(\omega)$:
\begin{equation}
S_\varphi(\omega)=4\frac{e^2}{\hbar^2}S_{U_1U_1}(\omega)\frac{\sin^2(\omega\tau/2)}{\omega^2}.
\end{equation}
We now determine the relation between the potential $U_1(t)$ and
the electrochemical potential $V_2(t)$, following the lines of
ref.\cite{Seelig01PRB64n245313}. In Fig. 1b, we have represented
$Q_1$ as the charge on the arm (a) capacitively coupled through
$C$ to the inner channel with a charge $Q_2$. The total charge on
the capacitance is the sum of an emitted charge and a screening
charge: $Q_j(\omega)=i
G_Q(1-e^{i\omega\tau})[V_j(\omega)-U_j(\omega)]/\omega$, with $G_Q
= e^2/h$, $j=1,2$ and $x(t)=\int x(\omega)e^{-i\omega
t}\mathrm{d}\omega$. From the charge neutrality $Q_1=-Q_2$ and the
$U(\omega)=U_2(\omega)-U_1(\omega)=Q_2(\omega)/C$, one gets
\begin{equation}
 G_{12}=\frac{dI_1(\omega)}{dV_2(\omega)}=\frac{G_Q(1-e^{i\omega\tau})}{2+iG_Q(1-e^{i\omega\tau})/(\omega
 C)}.
\end{equation}
\begin{figure}
\centerline{\includegraphics[angle=-90,width=8cm,keepaspectratio,clip]{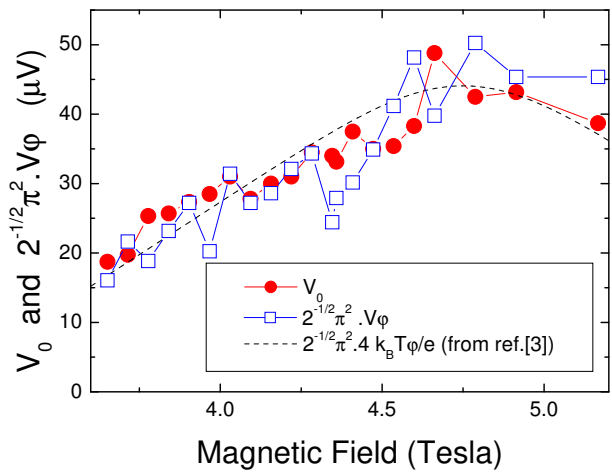}}
\caption{(color online) $V_0$ and $\pi^2V_\varphi/\sqrt{2}$ as a
function of the magnetic field. The dashed line is the general
behavior of $4k_BT_\varphi/e$ (right scale) measured in
ref.\cite{Roulleau0710.2806}, on the same
sample.}\label{expdecay.fig}
\end{figure}

This result is similar to the one obtained in
\cite{Seelig01PRB64n245313}, with differences in the charge
relaxation resistance: it arises both from the chirality of the
edge state ($R_Q/2$ instead of two $R_Q/2$'s in parallel) and from
the non "perfect" gate that forms the inner edge state (two
$R_Q/2$'s in series). The low frequency equivalent circuit (Fig.
1c) leads to $U_1=V_2/(C_Q/C+2)$ for dc biasing, $C_Q=G_Q\tau$.
Then the periodicity $V_0$ is related to $\tau$ by :
\begin{equation}
V_0=(2+1/\gamma)\frac{h}{e \tau}, \label{V_0.eq}
\end{equation}
with $\gamma=C/C_Q$. As $C\propto L$ and $C_Q\propto L/v_D$,
$\gamma$ should varies like $v_D$. Finally, the finite frequency
transconductance $G_{12}(\omega)$ allows to relate the potential
fluctuations to the electrochemical fluctuations:
\begin{equation}
S_{U_1U_1}(\omega)=\frac{1}{4}|1-G_{12}(\omega)i/\omega C|^2
S_{22}(\omega).
\end{equation}
We now consider the case of white partition noise $S_{22}=
2eR_QV_2\T_0(1-\T_0)$ or white thermal noise $S_{22}=
2\times4k_BR_QT$.
Finally, using Eqs. 5 and 8 with
$\tau/\tau_\varphi=<\delta\varphi^2>/2$, one finds:
\begin{equation}
\frac{1}{T_\varphi}= \frac{8 k_B}{\hbar}\,\tau {\cal I}(\gamma)
\;\mathrm{and}\; \frac{1}{V_\varphi}= \frac{2e}{\hbar}\,\tau {\cal
I}(\gamma)
\end{equation}
with,
$$
{\cal I}(\gamma)=\int_0^\infty
\frac{\sin^2(x)\gamma^2}{\sin^2(x)+2\gamma x\sin(2x)+4\gamma^2
x^2} dx.
$$

We see here that both the dephasing rate described by
$V_{\varphi}$ and $T_{\varphi}$ and the phase sensitivity to the
potential of the inner edge state described by $V_0$ are
proportional to the time of flight $\tau$. For $\gamma \ll 1$,
${\cal I}(\gamma)\sim \gamma \pi\sqrt{2}/8$. Then, using Eqs. 7
and 9,
\begin{equation}
V_\varphi^{-1}= \frac{\pi^2}{\sqrt{2}}\times V_0^{-1}.
\end{equation}

One can see, in Fig. 4, that our theory is in excellent agreement
with our measurements. Moreover, at first order, the ratio between
$V_0$ and $V_\varphi$ is independent of the coupling parameter
$\gamma$. It is not possible to conclude on the exact origin of
magnetic field variations of the coupling. It may arise from the
disorder which modifies the effective trajectory length and hence
the time of flight. The agreement between experiment and theory is
very good as far as $\gamma \lesssim 0.2$, which corresponds to
reasonable values of the capacitance and the drift velocity
\cite{Roulleau07PRB76n161309}. One can notice that
$\tau_\varphi^{-1}\propto \gamma$ when $\gamma\ll1$. Then,
increasing the coupling (or reducing the screening) increases
$\gamma$ and thus $1/\tau_\varphi$: the poor screening in the IQHE
is actually responsible for the limited coherence time in the edge
states.

To conclude, we have shown that the coherence length of the edge
states at filling factor 2 is limited by the Jonhson-Nyquist
noise. Changing the magnetic field makes it possible to modify the
coupling between the edge states and thus modifies the coherence
length. Our results are well described by a mean-field approach
that relates the phase randomization to the fluctuations of the
electrostatic potential in the interferometer arms.

The authors would like to thank Markus Buttiker for fruitfull
discussions.


\end{document}